\documentclass[10pt,a4paper,onecolumn]{article}
\usepackage{marginnote}
\usepackage{graphicx}
\usepackage{xcolor}
\usepackage{authblk,etoolbox}
\usepackage{titlesec}
\usepackage{calc}
\usepackage{tikz}
\usepackage{hyperref}
\hypersetup{colorlinks,breaklinks,
            urlcolor=[rgb]{0.0, 0.5, 1.0},
            linkcolor=[rgb]{0.0, 0.5, 1.0}}
\usepackage{caption}
\usepackage{tcolorbox}
\usepackage{amssymb,amsmath}
\usepackage{ifxetex,ifluatex}
\usepackage{seqsplit}
\usepackage{fixltx2e} 
\usepackage[
  backend=biber,
]{biblatex}

\newlength{\cslhangindent}
\newenvironment{cslreferences}%
  {\setlength{\parindent}{0pt}%
  \everypar{\setlength{\hangindent}{\cslhangindent}}\ignorespaces}%
  {\par}

\usepackage[top=3.5cm, bottom=3cm, right=1.5cm, left=1.0cm,
            headheight=2.2cm, reversemp, includemp, marginparwidth=4.5cm]{geometry}

\titleformat{\section}
  {\normalfont\sffamily\Large\bfseries}
  {}{0pt}{}
\titleformat{\subsection}
  {\normalfont\sffamily\large\bfseries}
  {}{0pt}{}
\titleformat{\subsubsection}
  {\normalfont\sffamily\bfseries}
  {}{0pt}{}
\titleformat*{\paragraph}
  {\sffamily\normalsize}

\usepackage{fancyhdr}
\makeatletter
\let\ps@plain\ps@fancy
\fancyheadoffset[L]{4.5cm}
\fancyfootoffset[L]{4.5cm}

\definecolor{linky}{rgb}{0.0, 0.5, 1.0}

\newtcolorbox{repobox}
   {colback=red, colframe=red!75!black,
     boxrule=0.5pt, arc=2pt, left=6pt, right=6pt, top=3pt, bottom=3pt}

\newcommand{\ExternalLink}{%
   \tikz[x=1.2ex, y=1.2ex, baseline=-0.05ex]{%
       \begin{scope}[x=1ex, y=1ex]
           \clip (-0.1,-0.1)
               --++ (-0, 1.2)
               --++ (0.6, 0)
               --++ (0, -0.6)
               --++ (0.6, 0)
               --++ (0, -1);
           \path[draw,
               line width = 0.5,
               rounded corners=0.5]
               (0,0) rectangle (1,1);
       \end{scope}
       \path[draw, line width = 0.5] (0.5, 0.5)
           -- (1, 1);
       \path[draw, line width = 0.5] (0.6, 1)
           -- (1, 1) -- (1, 0.6);
       }
   }

\patchcmd{\@maketitle}{center}{flushleft}{}{}
\patchcmd{\@maketitle}{center}{flushleft}{}{}
\patchcmd{\@maketitle}{\LARGE}{\LARGE\sffamily}{}{}
\def\maketitle{{%
  
  \AB@maketitle}}
\makeatletter
\renewcommand\AB@affilsepx{ \protect\Affilfont}
\renewcommand\AB@affilnote[1]{{\bfseries #1}\hspace{3pt}}
\makeatother

\renewcommand\Affilfont{\sffamily\small\mdseries}
\ifnum 0\ifxetex 1\fi\ifluatex 1\fi=0 
  \usepackage[T1]{fontenc}
  \usepackage[utf8]{inputenc}

\else 
  \ifxetex
    \usepackage{mathspec}
  \else
    \usepackage{fontspec}
  \fi
  \defaultfontfeatures{Ligatures=TeX,Scale=MatchLowercase}

\fi
\IfFileExists{upquote.sty}{\usepackage{upquote}}{}
\IfFileExists{microtype.sty}{%
\usepackage{microtype}
\UseMicrotypeSet[protrusion]{basicmath} 
}{}

\usepackage{hyperref}
\hypersetup{unicode=true,
            pdftitle={21cmFAST v3: A Python-integrated C code for generating 3D realizations of the cosmic 21cm signal.},
            pdfborder={0 0 0},
            breaklinks=true}
\usepackage{color}
\usepackage{fancyvrb}

\DefineVerbatimEnvironment{Highlighting}{Verbatim}{commandchars=\\\{\}}
\newenvironment{Shaded}{}{}

\newcommand{\CommentTok}[1]{\textcolor[rgb]{0.38,0.63,0.69}{\textit{#1}}}

\newcommand{\DecValTok}[1]{\textcolor[rgb]{0.25,0.63,0.44}{#1}}

\newcommand{\FloatTok}[1]{\textcolor[rgb]{0.25,0.63,0.44}{#1}}

\newcommand{\ImportTok}[1]{#1}

\newcommand{\NormalTok}[1]{#1}
\newcommand{\OperatorTok}[1]{\textcolor[rgb]{0.40,0.40,0.40}{#1}}

\newcommand{\StringTok}[1]{\textcolor[rgb]{0.25,0.44,0.63}{#1}}
\newcommand{\VariableTok}[1]{\textcolor[rgb]{0.10,0.09,0.49}{#1}}

\usepackage{longtable,booktabs}
\usepackage{graphicx,grffile}
\makeatletter
\def\maxwidth{\ifdim\Gin@nat@width>\linewidth\linewidth\else\Gin@nat@width\fi}
\def\maxheight{\ifdim\Gin@nat@height>\textheight\textheight\else\Gin@nat@height\fi}
\makeatother
\setkeys{Gin}{width=\maxwidth,height=\maxheight,keepaspectratio}
\IfFileExists{parskip.sty}{%
\usepackage{parskip}
}{
\setlength{\parindent}{0pt}
\setlength{\parskip}{6pt plus 2pt minus 1pt}
}
\providecommand{\tightlist}{%
  \setlength{\itemsep}{0pt}\setlength{\parskip}{0pt}}
\ifx\paragraph\undefined\else
\let\oldparagraph\paragraph
\renewcommand{\paragraph}[1]{\oldparagraph{#1}\mbox{}}
\fi
\ifx\subparagraph\undefined\else
\let\oldsubparagraph\subparagraph
\renewcommand{\subparagraph}[1]{\oldsubparagraph{#1}\mbox{}}
\fi

\title{21cmFAST v3: A Python-integrated C code for generating 3D
realizations of the cosmic 21cm signal.}

        \author[1]{Steven G. Murray}
          \author[2, 3]{Bradley Greig}
          \author[4]{Andrei Mesinger}
          \author[5]{Julian B. Muñoz}
          \author[4]{Yuxiang Qin}
          \author[4, 7]{Jaehong Park}
          \author[6]{Catherine A. Watkinson}
    
      \affil[1]{School of Earth and Space Exploration, Arizona State
University, Phoenix, USA}
      \affil[2]{ARC Centre of Excellence for All-Sky Astrophysics in 3
Dimensions (ASTRO 3D)}
      \affil[3]{School of Physics, University of Melbourne, Parkville,
VIC 3010, Australia}
      \affil[4]{Scuola Normale Superiore, Piazza dei Cavalieri 7, 56126
Pisa, Italy}
      \affil[5]{Department of Physics, Harvard University, 17 Oxford
St., Cambridge, MA, 02138, USA}
      \affil[6]{School of Physics and Astronomy, Queen Mary University
of London, G O Jones Building, 327 Mile End Road, London, E1 4NS, UK}
      \affil[7]{School of Physics, Korea Institute for Advanced Study,
85 Hoegiro, Dongdaemun-gu, Seoul, 02455, Republic of Korea}
  \date{\vspace{-5ex}}

\begin{document}
\maketitle

\marginpar{
  \sffamily\small

  {\bfseries DOI:} \href{https://doi.org/10.21105/joss.00850}{\color{linky}{10.21105/joss.00850}}

  \vspace{2mm}

  {\bfseries Software}
  \begin{itemize}
    \setlength\itemsep{0em}
    \item \href{https://github.com/openjournals/joss-reviews/issues/54}{\color{linky}{Review}} \ExternalLink
    \item \href{https://github.com/21cmfast/21cmFAST}{\color{linky}{Repository}} \ExternalLink
    \item \href{http://dx.doi.org/10.21105/zenodo.1400822}{\color{linky}{Archive}} \ExternalLink
  \end{itemize}

  \vspace{2mm}

  {\bfseries Submitted:} 20 July 2020\\
  {\bfseries Published:} 22 October 2020

  \vspace{2mm}
  {\bfseries License}\\
  Authors of papers retain copyright and release the work under a Creative Commons Attribution 4.0 International License (\href{http://creativecommons.org/licenses/by/4.0/}{\color{linky}{CC-BY}}).
}

\hypertarget{summary}{%
\section{Summary}\label{summary}}

The field of 21-cm cosmology -- in which the hyperfine spectral line of
neutral hydrogen (appearing at the rest-frame wavelength of 21 cm) is
mapped over large swathes of the Universe's history -- has developed
radically over the last decade. The promise of the field is to
revolutionize our knowledge of the first stars, galaxies, and black
holes through the timing and patterns they imprint on the cosmic 21-cm
signal. In order to interpret the eventual observational data, a range
of physical models have been developed -- from simple analytic models of
the global history of hydrogen reionization, through to fully
hydrodynamical simulations of the 3D evolution of the brightness
temperature of the spectral line. Between these extremes lies an
especially versatile middle-ground: fast semi-numerical models that
approximate the full 3D evolution of the relevant fields: density,
velocity, temperature, ionization, and radiation (Lyman-alpha, neutral
hydrogen 21-cm, etc.). These have the advantage of being comparable to
the full first-principles hydrodynamic simulations, but significantly
quicker to run; so much so that they can be used to produce thousands of
realizations on scales comparable to those observable by upcoming
low-frequency radio telescopes, in order to explore the very wide
parameter space that still remains consistent with the data.

Amongst practitioners in the field of 21-cm cosmology, the
\texttt{21cmFAST} program has become the \emph{de facto} standard for
such semi-numerical simulators. \texttt{21cmFAST} (Mesinger and
Furlanetto 2007; Mesinger, Furlanetto, and Cen 2011) is a
high-performance C code that uses the excursion set formalism
(Furlanetto, Zaldarriaga, and Hernquist 2004) to identify regions of
ionized hydrogen atop a cosmological density field evolved using first-
or second-order Lagrangian perturbation theory (Zel'Dovich 1970;
Scoccimarro and Sheth 2002), tracking the thermal and ionization state
of the intergalactic medium, and computing X-ray, soft UV and ionizing
UV cosmic radiation fields based on parametrized galaxy models. For
example, the following figure contains slices of lightcones (3D fields
in which one axis corresponds to both spatial \emph{and} temporal
evolution) for the various component fields produced by
\texttt{21cmFAST}.

\begin{figure}
\centering
\includegraphics[width=\textwidth,height=4.6875in]{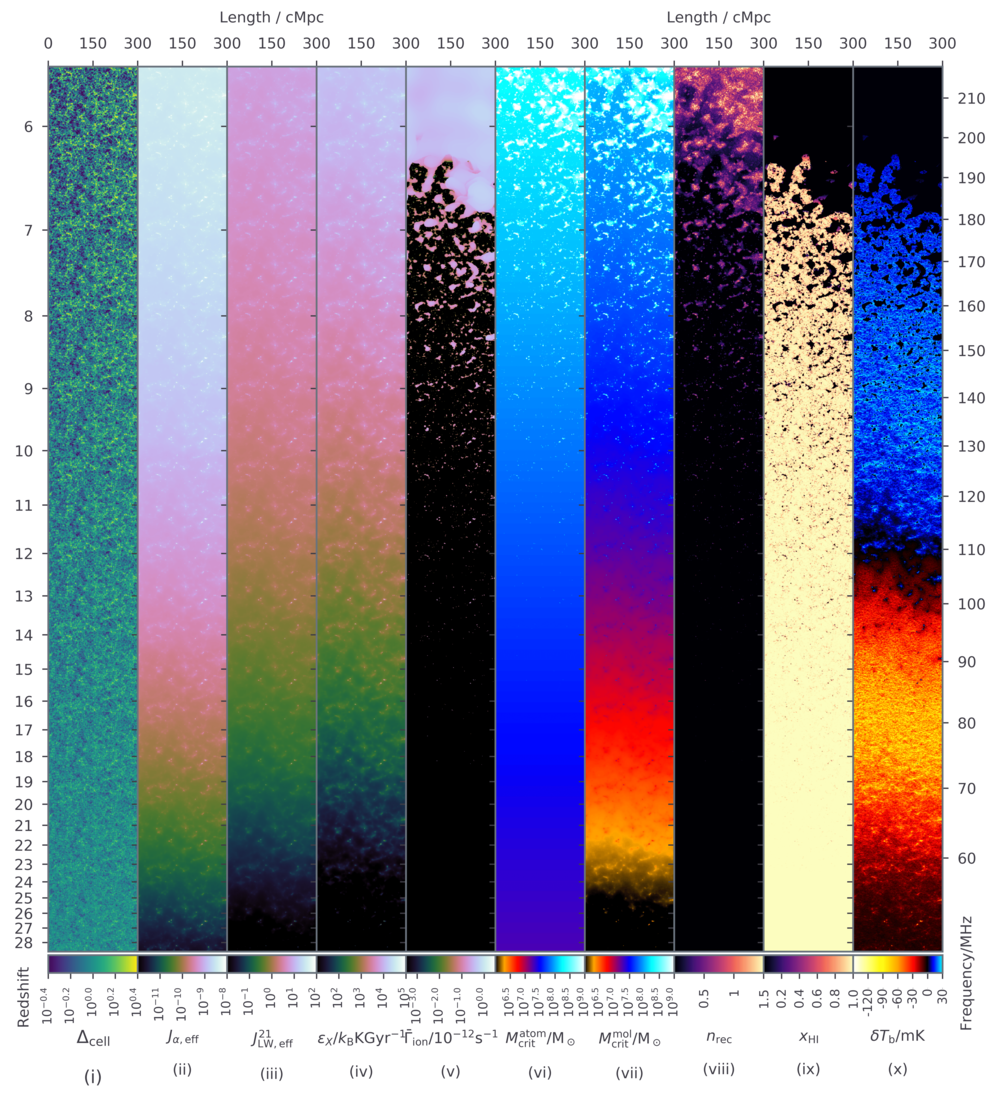}
\caption{Sample of Component Fields output by 21cmFAST. Cosmic evolution
occurs from bottom to top. From left to right, quantities shown are: (i)
dark matter overdensity field; (ii) Lyman-alpha flux; (iii) Lyman-Werner
flux; (iv) X-ray heating rate; (v) locally-averaged UVB; (vi) critical
halo mass for star formation in Atomically Cooled Galaxies; (vii)
critical halo mass for star formation in Molecularly Cooled Galaxies;
(viii) cumulative number of recombinations per baryon; (ix) neutral
hydrogen fraction; and (x) the 21cm brightness temperature. A
high-resolution version of this figure is available at
\url{http://homepage.sns.it/mesinger/Media/lightcones_minihalo.png}}
\end{figure}

However, \texttt{21cmFAST} is a highly specialized code, and its
implementation has been quite specific and relatively inflexible. This
inflexibility makes it difficult to modify the behaviour of the code
without detailed knowledge of the full system, or disrupting its
workings. This lack of modularity within the code has led to widespread
code ``branching'' as researchers hack new physical features of interest
into the C code; the lack of a streamlined API has led derivative codes
which run multiple realizations of \texttt{21cmFAST} simulations (such
as the Monte Carlo simulator, \texttt{21CMMC}, Greig and Mesinger 2015)
to re-write large portions of the code in order to serve their purpose.
It is thus of critical importance, as the field moves forward in its
understanding -- and the range and scale of physical models of interest
continues to increase -- to reformulate the \texttt{21cmFAST} code in
order to provide a fast, modular, well-documented, well-tested, stable
simulator for the community.

\hypertarget{features-of-21cmfast-v3}{%
\section{Features of 21cmFAST v3}\label{features-of-21cmfast-v3}}

This paper presents \texttt{21cmFAST} v3+, which is formulated to follow
these essential guiding principles. While keeping the same core
functionality of previous versions of \texttt{21cmFAST}, it has been
fully integrated into a Python package, with a simple and intuitive
interface, and a great deal more flexibility. At a higher level, in
order to maintain best practices, a community of users and developers
has coalesced into a formal collaboration which maintains the project
via a Github organization. This allows the code to be consistently
monitored for quality, maintaining high test coverage, stylistic
integrity, dependable release strategies and versioning, and peer code
review. It also provides a single point-of-reference for the community
to obtain the code, report bugs and request new features (or get
involved in development).

A significant part of the work of moving to a Python interface has been
the development of a robust series of underlying Python structures which
handle the passing of data between Python and C via the \texttt{CFFI}
library. This foundational work provides a platform for future versions
to extend the scientific capabilities of the underlying simulation code.
The primary \emph{new} usability features of \texttt{21cmFAST} v3+ are:

\begin{itemize}
\tightlist
\item
  Convenient (Python) data objects which simplify access to and
  processing of the various fields that form the brightness temperature.
\item
  Enhancement of modularity: the underlying C functions for each step of
  the simulation have been de-coupled, so that arbitrary functionality
  can be injected into the process.
\item
  Conversion of most global parameters to local structs to enable this
  modularity, and also to obviate the requirement to re-compile in order
  to change parameters.
\item
  Simple \texttt{pip}-based installation.
\item
  Robust on-disk caching/writing of data, both for efficiency and
  simplified reading of previously processed data (using HDF5).
\item
  Simple high-level API to generate either coeval cubes (purely spatial
  3D fields defined at a particular time) or full lightcone data
  (i.e.~those coeval cubes interpolated over cosmic time, mimicking
  actual observations).
\item
  Improved exception handling and debugging.
\item
  Convenient plotting routines.
\item
  Simple configuration management, and also more intuitive management
  for the remaining C global variables.
\item
  Comprehensive API documentation and tutorials.
\item
  Comprehensive test suite (and continuous integration).
\item
  Strict semantic versioning\footnote{\url{https://semver.org}}.
\end{itemize}

While in v3 we have focused on the establishment of a stable and
extendable infrastructure, we have also incorporated several new
scientific features, appearing in separate papers:

\begin{itemize}
\tightlist
\item
  Generate transfer functions using the \texttt{CLASS} Boltzmann code
  (Lesgourgues 2011).
\item
  Simulate the effects of relative velocities between dark matter and
  Baryons (Muñoz 2019b, 2019a).
\item
  Correction for non-conservation of ionizing photons (Park, Greig et
  al., \emph{in prep}).
\item
  Include molecularly cooled galaxies with distinct properties (Qin,
  Mesinger, et al. 2020)
\item
  Calculate rest-frame UV luminosity functions based on parametrized
  galaxy models.
\end{itemize}

\texttt{21cmFAST} is still in very active development. Amongst further
usability and performance improvements, future versions will see several
new physical models implemented, including milli-charged dark matter
models (Muñoz, Dvorkin, and Loeb 2018) and forward-modelled CMB
auxiliary data (Qin, Poulin, et al. 2020).

In addition, \texttt{21cmFAST} will be incorporated into large-scale
inference codes, such as \texttt{21CMMC}, and is being used to create
large data-sets for inference via machine learning. We hope that with
this new framework, \texttt{21cmFAST} will remain an important component
of 21-cm cosmology for years to come.

\hypertarget{examples}{%
\section{Examples}\label{examples}}

\texttt{21cmFAST} supports installation using \texttt{conda}, which
means installation is as simple as typing
\texttt{conda\ install\ -c\ conda-forge\ 21cmFAST}. The following
example can then be run in a Python interpreter.

In-depth examples can be found in the official documentation. As an
example of the simplicity with which a full lightcone may be produced
with the new \texttt{21cmFAST} v3, the following may be run in a Python
interpreter (or Jupyter notebook):

\begin{Shaded}
\begin{Highlighting}[]
\ImportTok{import}\NormalTok{ py21cmfast }\ImportTok{as}\NormalTok{ p21c}

\NormalTok{lightcone }\OperatorTok{=}\NormalTok{ p21c.run\_lightcone(}
\NormalTok{    redshift}\OperatorTok{=}\FloatTok{6.0}\NormalTok{,              }\CommentTok{\# Minimum redshift of lightcone}
\NormalTok{    max\_redshift}\OperatorTok{=}\FloatTok{30.0}\NormalTok{,}
\NormalTok{    user\_params}\OperatorTok{=}\NormalTok{\{}
        \StringTok{"HII\_DIM"}\NormalTok{: }\DecValTok{150}\NormalTok{,        }\CommentTok{\# N cells along side in output cube}
        \StringTok{"DIM"}\NormalTok{: }\DecValTok{400}\NormalTok{,            }\CommentTok{\# Original high{-}res cell number}
        \StringTok{"BOX\_LEN"}\NormalTok{: }\DecValTok{300}\NormalTok{,        }\CommentTok{\# Size of the simulation in Mpc}
\NormalTok{    \},}
\NormalTok{    flag\_options}\OperatorTok{=}\NormalTok{\{}
        \StringTok{"USE\_TS\_FLUCT"}\NormalTok{: }\VariableTok{True}\NormalTok{,  }\CommentTok{\# Don\textquotesingle{}t assume saturated spin temp}
        \StringTok{"INHOMO\_RECO"}\NormalTok{: }\VariableTok{True}\NormalTok{,   }\CommentTok{\# Use inhomogeneous recombinations}
\NormalTok{    \},}
\NormalTok{    lightcone\_quantities}\OperatorTok{=}\NormalTok{(     }\CommentTok{\# Components to store as lightcones}
        \StringTok{"brightness\_temp"}\NormalTok{,}
        \StringTok{"xH\_box"}\NormalTok{,}
        \StringTok{"density"}
\NormalTok{    ),}
\NormalTok{    global\_quantities}\OperatorTok{=}\NormalTok{(        }\CommentTok{\# Components to store as mean}
        \StringTok{"xH\_box"}\NormalTok{,              }\CommentTok{\# values per redshift}
        \StringTok{"brightness\_temp"}
\NormalTok{    ),}
\NormalTok{)}

\CommentTok{\# Save to a unique filename hashing all input parameters}
\NormalTok{lightcone.save()}

\CommentTok{\# Make a lightcone sliceplot}
\NormalTok{p21c.plotting.lightcone\_sliceplot(lightcone, }\StringTok{"brightness\_temp"}\NormalTok{)}
\end{Highlighting}
\end{Shaded}

\begin{figure}
\centering
\includegraphics[width=\textwidth,height=3.125in]{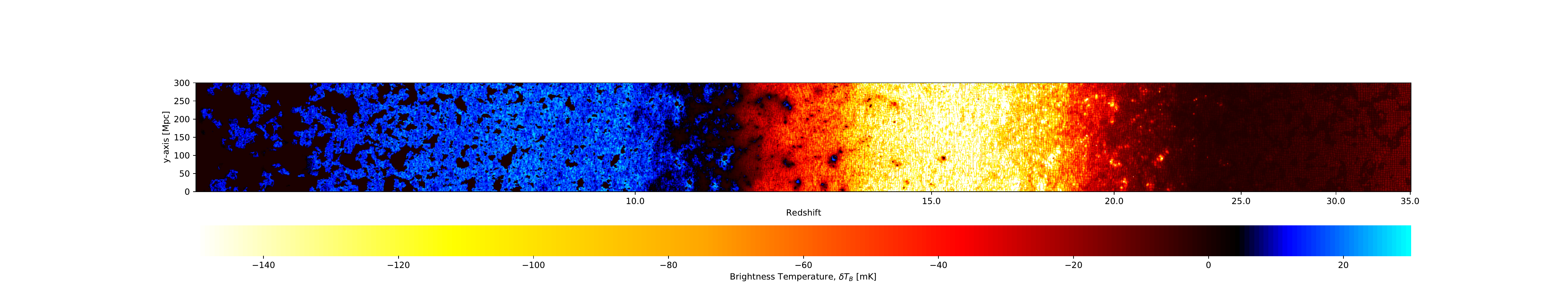}
\caption{Brightness temperature lightcone produced by the example code
in this paper.}
\end{figure}

\begin{Shaded}
\begin{Highlighting}[]
\CommentTok{\# Plot a global quantity}
\NormalTok{p21c.plotting.plot\_global\_history(lightcone, }\StringTok{"xH"}\NormalTok{)}
\end{Highlighting}
\end{Shaded}

\begin{figure}
\centering
\includegraphics[width=\textwidth,height=3.125in]{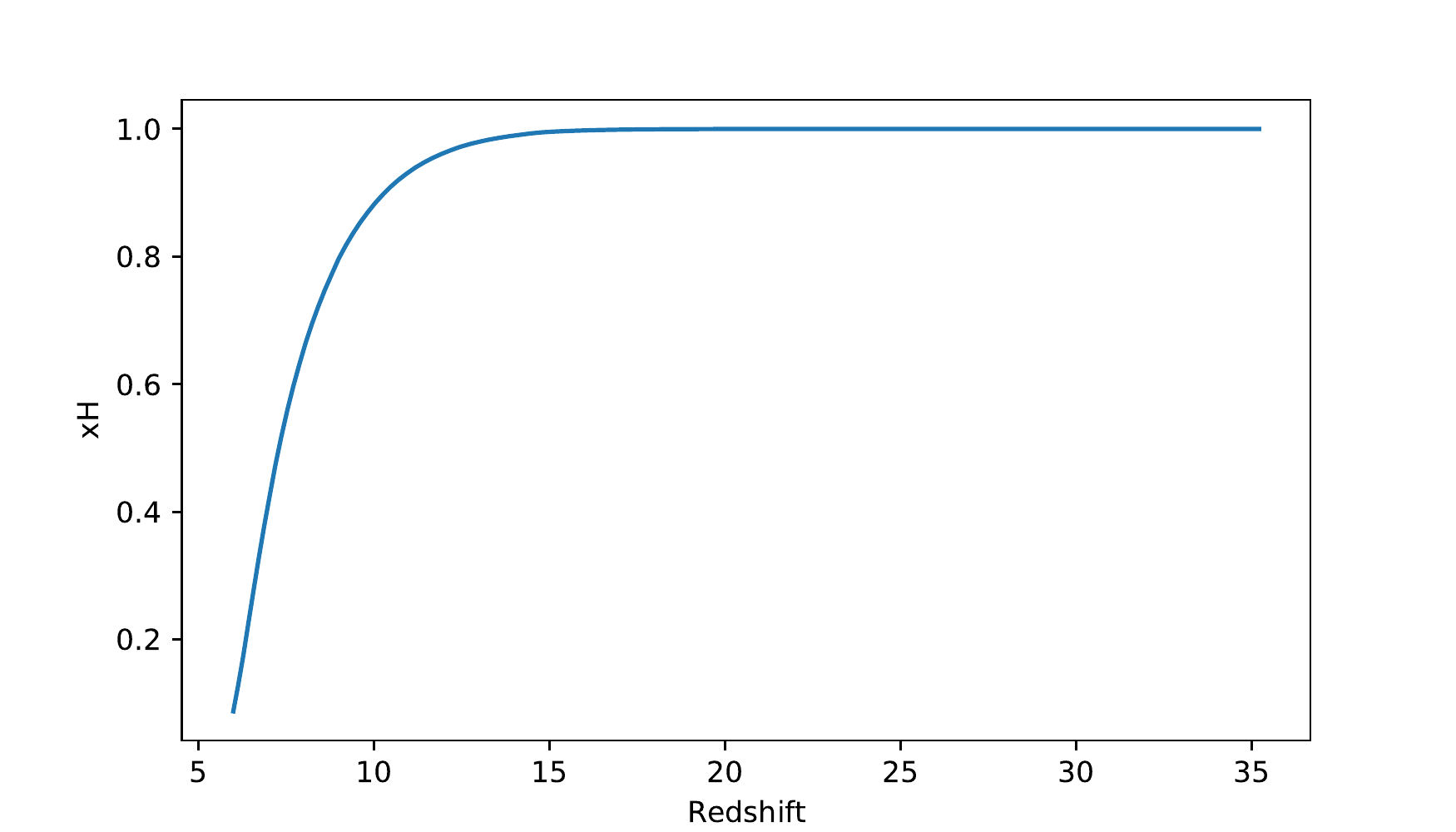}
\caption{Globally volume-averaged hydrogen neutral fraction produced by
the example code in this paper.}
\end{figure}

\hypertarget{performance}{%
\section{Performance}\label{performance}}

Despite being a Python code, \texttt{21cmFAST} v3 does not diminish the
performance of previous pure-C versions. It utilises \texttt{CFFI} to
provide the interface to the C-code through Python, which is managed by
some custom Python classes that oversee the construction and memory
allocation of each C \texttt{struct}.

OpenMP parallelization is enabled within the C-code, providing excellent
speed-up for large simulations when performed on high-performance
machines.

A simple performance comparison between v3 and v2.1 (the last pure-C
version), running a light-cone simulation over a redshift range between
35 and 5 (92 snapshots) with spin temperature fluctuations
(\texttt{USE\_TS\_FLUCT}), inhomogeneous recombinations
(\texttt{INHOMO\_RECO}), FFTW Wisdoms (\texttt{USE\_FFTW\_WISDOM}) and
interpolation tables (\texttt{USE\_INTERPOLATION\_TABLES}), with a
resolution of \texttt{HII\_DIM=250} cells, and \texttt{DIM=1000} cells
for the initial conditions, on an Intel(R) Xeon(R) CPU (E5-4657L v2 @
2.40GHz) with 16 shared-memory cores, reveals that a clock time of
7.63(12.63) hours and a maximum RAM of 224(105) gigabytes are needed for
v3(v2.1).

Note that while a full light-cone simulation can be expensive to
perform, it only takes 2-3min to calculate a Coeval box (excluding the
initial conditions). For instance, the aforementioned timing for v3
includes 80 minutes to generate the initial condition, which also
dominates the maximum RAM required, with an additional \textasciitilde4
minutes per snapshot to calculate all required fields of perturbation,
ionization, spin temperature and brightness temperature.

To guide the user, we list some performance benchmarks for variations on
this simulation, run with \texttt{21cmFAST} v3.0.2. Note that these
benchmarks are subject to change as new minor versions are delivered; in
particular, operational modes that reduce maximum memory consumption are
planned for the near future.

\begin{longtable}[]{@{}lll@{}}
\toprule
Variation & Time (hr) & Memory (GB)\tabularnewline
\midrule
\endhead
Reference & 7.63 & 224\tabularnewline
Single Core & 14.77 & 224\tabularnewline
4 Shared-memory Cores & 7.42 & 224\tabularnewline
64 Shared-memory Cores & 9.60 & 224\tabularnewline
Higher Resolution (HII\_DIM=500, DIM=2000) & 68.37 & 1790\tabularnewline
Lower Resolution (HII\_DIM=125, DIM=500) & 0.68 & 28\tabularnewline
No Spin Temperature & 4.50 & 224\tabularnewline
Use Mini-Halos & 11.57 & 233\tabularnewline
No FFTW Wisdoms & 7.33 & 224\tabularnewline
\bottomrule
\end{longtable}

At this time, the \texttt{21cmFAST} team suggests using 4 or fewer
shared-memory cores. However, it is worth noting that as performance
does vary on different machines, users are recommended to calculate
their own scalability.

\hypertarget{acknowledgements}{%
\section{Acknowledgements}\label{acknowledgements}}

This work was supported in part by the European Research Council (ERC)
under the European Union's Horizon 2020 research and innovation
programme (AIDA -- \#638809). The results presented here reflect the
authors' views; the ERC is not responsible for their use. JBM was
partially supported by NSF grant AST-1813694. Parts of this research
were supported by the European Research Council under ERC grant number
638743-FIRSTDAWN. Parts of this research were supported by the
Australian Research Council Centre of Excellence for All Sky
Astrophysics in 3 Dimensions (ASTRO 3D), through project number
CE170100013. JP was supported in part by a KIAS individual Grant
(PG078701) at Korea Institute for Advanced Study.

\hypertarget{references}{%
\section*{References}\label{references}}
\addcontentsline{toc}{section}{References}

\hypertarget{refs}{}
\begin{cslreferences}
\leavevmode\hypertarget{ref-furlanetto2004}{}%
Furlanetto, Steven R., Matias Zaldarriaga, and Lars Hernquist. 2004.
``The Growth of H II Regions During Reionization.'' \emph{The
Astrophysical Journal} 613 (1): 1--15.
\url{https://doi.org/10.1086/423025}.

\leavevmode\hypertarget{ref-greig2015}{}%
Greig, Bradley, and Andrei Mesinger. 2015. ``21CMMC: an MCMC analysis
tool enabling astrophysical parameter studies of the cosmic 21 cm
signal.'' \emph{Monthly Notices of the Royal Astronomical Society} 449
(4): 4246--63. \url{https://doi.org/10.1093/mnras/stv571}.

\leavevmode\hypertarget{ref-Lesgourgues2011}{}%
Lesgourgues, Julien. 2011. ``The Cosmic Linear Anisotropy Solving System
(CLASS) I: Overview.'' \emph{arXiv E-Prints}, April, arXiv:1104.2932.
\url{http://arxiv.org/abs/1104.2932}.

\leavevmode\hypertarget{ref-mesinger2007}{}%
Mesinger, Andrei, and Steven Furlanetto. 2007. ``Efficient Simulations
of Early Structure Formation and Reionization.'' \emph{The Astrophysical
Journal} 669 (2): 663--75. \url{https://doi.org/10.1086/521806}.

\leavevmode\hypertarget{ref-mesinger2010}{}%
Mesinger, Andrei, Steven Furlanetto, and Renyue Cen. 2011. ``21CMFAST: a
fast, seminumerical simulation of the high-redshift 21-cm signal.''
\emph{Monthly Notices of the Royal Astronomical Society} 411 (2):
955--72. \url{https://doi.org/10.1111/j.1365-2966.2010.17731.x}.

\leavevmode\hypertarget{ref-munoz2019b}{}%
Muñoz, Julian B. 2019a. ``Robust velocity-induced acoustic oscillations
at cosmic dawn.'' \emph{Physics Review D} 100 (6): 063538.
\url{https://doi.org/10.1103/PhysRevD.100.063538}.

\leavevmode\hypertarget{ref-munoz2019a}{}%
---------. 2019b. ``Standard Ruler at Cosmic Dawn.'' \emph{Physics
Review Letters} 123 (13): 131301.
\url{https://doi.org/10.1103/PhysRevLett.123.131301}.

\leavevmode\hypertarget{ref-Munoz2018}{}%
Muñoz, Julian B., Cora Dvorkin, and Abraham Loeb. 2018. ``21-cm
Fluctuations from Charged Dark Matter.'' \emph{Physics Review Letters}
121 (12): 121301. \url{https://doi.org/10.1103/PhysRevLett.121.121301}.

\leavevmode\hypertarget{ref-qin2020}{}%
Qin, Yuxiang, Andrei Mesinger, Jaehong Park, Bradley Greig, and Julian
B. Muñoz. 2020. ``A tale of two sites - I. Inferring the properties of
minihalo-hosted galaxies from current observations.'' \emph{Monthly
Notices of the Royal Astronomical Society} 495 (1): 123--40.
\url{https://doi.org/10.1093/mnras/staa1131}.

\leavevmode\hypertarget{ref-qin2020a}{}%
Qin, Yuxiang, Vivian Poulin, Andrei Mesinger, Bradley Greig, Steven
Murray, and Jaehong Park. 2020. ``Reionization inference from the CMB
optical depth and E-mode polarization power spectra.'' \emph{Monthly
Notices of the Royal Astronomical Society} 499 (1): 550--58.
\url{https://doi.org/10.1093/mnras/staa2797}.

\leavevmode\hypertarget{ref-scoccimarro2002}{}%
Scoccimarro, Román, and Ravi K. Sheth. 2002. ``PTHALOS: a fast method
for generating mock galaxy distributions.'' \emph{Monthly Notices of the
Royal Astronomical Society} 329 (3): 629--40.
\url{https://doi.org/10.1046/j.1365-8711.2002.04999.x}.

\leavevmode\hypertarget{ref-zeldovich1970}{}%
Zel'Dovich, Y. B. 1970. ``Gravitational instability: an approximate
theory for large density perturbations.'' \emph{Astronomy and
Astrophysics} 500 (March): 13--18.
\end{cslreferences}

\end{document}